\documentclass[article]{JHEP}
\usepackage{amsmath,epsfig}
\usepackage{amssymb,amsfonts}
\usepackage{feynmf}
\usepackage{epsfig}
\usepackage[vcentermath,enableskew]{youngtab}
\relax
\renewcommand{\theequation}{\arabic{section}.\arabic{equation}}
\def\hre#1#2{\href{http://arxiv.org/abs/#1/#2}{[ArXiv:#1/#2]}}
\def\hspi#1#2{\href{http://www.slac.stanford.edu/spires/find/hep/www?irn=#1}{#2}}
\def\be{\begin{equation}}
\def\ee{\end{equation}}
\def\bea{\begin{eqnarray}}
\def\eea{\end{eqnarray}}
\def\be{\begin{equation}}
\def\ee{\end{equation}}
\def\eqp{\;\;.}

\def\pa{\partial}

\newcommand\fverb{\setbox\pippobox=\hbox\bgroup\verb}
\newcommand\fverbdo{\egroup\medskip\noindent%
                        \fbox{\unhbox\pippobox}\ }
\newcommand\fverbit{\egroup\item[\fbox{\unhbox\pippobox}]}
\newbox\pippobox

\def\e{\epsilon}

\def\m{\mu}
\def\n{\nu}

\def\s{\sigma}

\def\sp{\;\;\;,\;\;\;}

\def\a{\alpha}
\def\b{\beta}

\def\l{\lambda}

\def\ba{\begin{eqnarray}}
\def\ea{\end{eqnarray}}

\def\sla{\raise.15ex\hbox{$/$}\kern-.57em}
\title{\Huge Product CFTs, gravitational cloning, massive gravitons and the space of gravitational duals}
\author{{\Large Elias Kiritsis}{\Large $^{1,2}$}\\
~\\
~\\
$^1~$CPHT, UMR du CNRS 7644, Ecole Polytechnique,\\
91128, Palaiseau, FRANCE\\
$~$\\
$^2~$Department of Physics, University of Crete\\
71003 Heraklion, GREECE}

\preprint{\hepth{0608088}\\
CPHT-RR021.0406}

\abstract{The question of graviton cloning in the context of the bulk/boundary correspondence is considered.
It is  shown that multi-graviton theories can be obtained from products of large-N CFTs.
No more than one interacting massless graviton is possible. There can be however, many interacting massive gravitons.
This is achieved by coupling CFTs via multi-trace marginal or relevant perturbations.
The geometrical structure of the gravitational duals of such theories is that of product manifolds with their boundaries identified.
The calculational formalism is described and the  interpretation of such theories is discussed.}

\begin{document}



\section{Introduction}

The question of the non-triviality of theories with multiple or massive gravitons,
has been asked several  times in the past, \cite{deser,henneaux}.
In string theory, it has been argued \cite{bachas}, using the idea of topological cloning,
that in standard asymptotically flat vacua, the presence of multiple
 massless gravitons is only possible if the associated string theories do not interact.

With the advent of AdS/CFT correspondance, \cite{malda} and its string theory/gauge-theory generalizations,
we have in our hands information on non-asymptotically flat vacua of string theory.
It is therefore interesting to pose the question of non-trivial graviton cloning for
asymptotically AdS vacua of string theory.
This is the question that will be investigated in this paper. In the process we will uncover
a rich new structure in the
space of bulk (gravitational) duals of large-N gauge theories.

It has been expected that every large-N theory is dual to a string theory on a given
background space-time. We will see that the space of such space-times contains product spaces
with a common boundary.
The gravitational physics of such spaces is defined via coupled boundary conditions at the common boundary.
Although we mostly discuss CFTs in this paper, non-conformal theories are also considered, and it is obvious that the framework generalizes
directly.

Such product space-times arise, when we couple two large-N CFTs via perturbations that are products
 of operators belonging to both theories. Double-trace perturbations in a single CFT
and associated string theory, have already been  considered  in \cite{nonlocal}.
We will investigate such product theories in this paper and we will show that they are associated to massive gravitons with
 transparent boundary conditions. These were described in a somewhat different setting in \cite{porrati}.
The way massive gravitons appear is straightforward: one of the two conserved stress tensor of the unperturbed
pair of CFTs ceases to be conserved once we turn-on the coupled perturbation.
This will give rise to a massive graviton in an AdS space (or deformations thereof).\footnote{Of course string theory around flat space contains, an infinite number of massive spin-two
excitations. These have string scale masses, and arise at higher level than the massless one.
They form, among others, the Regge trajectory of the massless graviton.
In particular they are states, that have always masses at or above the
string scale and they can never become light. We are not interested in such massive graviton states
in this paper.
Our primary interest are massless gravitons, or massive gravitons
but with masses that are not necessarily at the string scale.}

\section{Massive and massless gravitons in AdS/CFT}

It is well known that the presence of a massless graviton in the bulk theory
dual to a boundary CFT, is intimately related to the presence of translation invariance
and energy conservation.
Indeed, to see this, we couple a source $h_{\m\n}$ to the stress tensor of the Field Theory
\be
e^{W_{eff}(h)}\equiv \int e^{-S_*+\int d^4x ~h_{\m\n}T^{\m\n}}
\label{1}
\ee
We now perform an infinitesimal diffeomorphism $x^{\m}\to x^{\m}+\e^{\m}$, under which the
action by definition transforms as
\be
S_*\to S_*-2\int d^4 x~\pa_{\m}\e_{\n}T^{\m\n}
\label{2}
\ee
If translations are symmetries, then the relation above gives energy-momentum conservation
\be
\pa_{\m}T^{\m\n}=0
\label{4}
\ee
The effective action is therefore invariant under diffeomorphisms acting on the metric
\be
W_{eff}(h_{\m\n}+\pa_{\m}\e_{\n}+\pa_{\n}\e_{\m})=W_{eff}(h_{\m\n})
\label{3}\ee
As usual in large-N conformal theories, the graviton Regge trajectory is lifted to a five-dimensional
graviton field. The ensuing diffeomorphism invariance, translates into the masslessness of the
five-dimensional graviton. This statement remains true in the presence of extra dimensions
 associated with other global symmetries of the
theory.

 To conclude, the energy-momentum conservation of the boundary CFT is inextricably related
to bulk diffeomorphism invariance and the masslessness of the graviton.
\footnote{The five-dimensional masslessness of the bulk graviton should not be confused with the non-zero
four-dimensional (background-dependent) masses the spin-two glueballs acquire in a given bulk vacuum
(associated with a concrete background solution).}

We may consider however the presence of other symmetric two-index traceless operators $\tilde T_{\m\n}$.
Such operators, generically exist in the spectrum of CFTs, and {\it are not} generically conserved,
\be
\pa_{\m}\tilde T^{\m\n}=J^{\n}\sp \pa^{\m}J_{\m}=\Phi
\label{5}\ee
If the dimension of $\tilde T_{\m\n}$ is $\Delta$, the dimension
of the operator $J^{\m}$ is $\Delta+1$ and that of $\Phi$, $\Delta+2$.
The operator $\tilde T_{\m\n}$ should correspond to another spin-two bulk field.
Now however, this field combines with the (massive) bulk vector
associated to the $J^{\m}$ and $\Phi$ operators
to form a massive graviton multiplet.

A CFT has an infinite number of spin-two gauge-invariant operators,
that give rise to spin-two fields in the bulk theory.
An example in ${\cal N}=4$ SYM theory consists of the series of operators
$Tr[\Phi^{I_1}\cdots \Phi^{I_{n-1}} D_{\m}D_{\n}\Phi^{I_n}]$.
Such operators are non-BPS and therefore have large scaling dimensions and the associated glueballs
string scale  masses.
Generically, such fields are associated to spin-two states with
string scale masses. In some cases however, they may correspond to low-lying gravitons.

In \cite{porrati}, the group theoretic analysis of conformal representations
associated with massive gravitons in $AdS_4$ was performed in detail. It was shown that the graviton
acquires a mass, if special transparent boundary conditions are chosen for scalars sourcing the stress
tensor of the CFT. We will see in the sequel that this observation is explained and generalized by our
results.

\section{Absence of massless interacting gravitons}

We will first examine and exclude the case of massless interacting gravitons.
Consider a theory with two traceless and conserved spin-two operators.
After diagonalizing their two-point functions we may
denote  the orthogonal set by $T_1^{\m\n}$ and $T_2^{\m\n}$.
They generate two commuting  conformal algebras.

The strategy is to diagonalize
the spectrum of operators under these algebras and
factorize the theory as a direct product of two sub-theories that are non-interacting.
If this is achieved, then we have obtained two theories each living independently
on $AdS_5\times X_1$ and $AdS_5\times X_2$. Since the correlators factorize, the theories
and therefore their gravitons are non-interacting.

There are caveats to the argument above, and they have been investigated in 2d CFTs which are
in general far better understood compared to their relatives in higher dimensions.
The issue of factorization of CFTs given two commuting stress-tenors has been exploited in two dimensions
in order to construct, by factorization, new CFTs from known ones.
Early examples go back to \cite{bh}, while this is the philosophy of the coset construction \cite{gko}.
The procedure has been described in full generality in \cite{ki} and has been applied to affine CFTs \cite{av}
in order to construct large classes of new CFTs ,
with generically irrational central charge. The general formalism is reviewed in
\cite{av-review}.
Moreover, there is also the exotic case of indecomposable representations, giving rise to
logarithmic CFTs in two dimensions, \cite{logarithmic}.
We will not consider this option further in this paper.

The upshot of these studies is that a theory with two commuting stress-tensors (that sum up to the total stress tensor)
is a (discrete) projection of a direct product of
 two non-interacting CFTs, $CFT_1\times CFT_2$, each described by the associated stress-tensor.
Therefore, the correlators of the CFT factorize into sums of product correlators\footnote{The factorization
ansatz was first described for the G/H theories in \cite{mrd}. It has been formalized and solved in \cite{ho}.}.

The analogous statement in the dual gravitational theory is that we must consider two independent
$AdS_{d+1}\times X_1$ and $AdS_{d+1}\times X_2$ geometries dual to $CFT_{1,2}$ respectively.
The two gravitons are massless and the two theories are essentially non-interacting.
Allowed operators are however discretely correlated.

\section{Interacting product CFTs in four dimensions\label{inter}}

We will now study a product of two four-dimensional CFTs $CFT_1\times CFT_2$ coupled together by a marginal (or relevant)
double-trace perturbation of the form $O_1O_2$ where $O_1$ is a scalar single-trace, gauge-invariant operator in
$CFT_1$ and $O_2$ is a scalar single-trace gauge-invariant operator in
$CFT_2$. We normalize $O_i\sim Tr[\cdots]$ so that their two-point functions  are normalized to one.
We also take $N_{1,2}\to \infty$ with their ratio fixed.

The general implementation of multiple trace perturbations in the context of AdS/CFT was
described in \cite{witten}-\cite{klebanov} following the original work of \cite{nonlocal}.

Consider a scalar operator of scaling dimension $\Delta$ in CFT$_1$, $O_{\Delta}$, and another of dimension $4-\Delta$
in CFT$_2$, $\tilde O_{4-\Delta}$.
Perturbing the product theory, $CFT_1\times CFT_2$, with
\be
\delta S=g\int d^4x~O_{\Delta}(x)\tilde O_{4-\Delta}(x)
\label{9}
\ee
 we will obtain a line of fixed points according to our analysis in appendix \ref{A} if $\Delta\not=2$.
 Otherwise, this is a marginally relevant perturbation.
The implementation of this perturbation in the dual theory, will start with the two
geometries of the unperturbed CFTs, $CFT_1\to AdS_5\times X_1$, $CFT_2\to AdS_5\times X_2$
where the compact internal spaces $X_{1,2}$ are not necessarily the same, as the two original CFTs need not be the same.
In particular they share the same set of boundary four-dimensional coordinates
 as the two CFTs live on the same four-dimensional manifold. However, the radial
  holographic coordinate as well as the coordinates of the spaces $X_{1,2}$ are distinct.

When the perturbation (\ref{9}) does not break conformal invariance, turning it on, does not modify the
$AdS_5$ parts of the metric. As we will see, it does not change the compact parts either, as the
associated bulk fields are trivial in the background solution.
In the more general case of relevant perturbations, the background solutions will get modified.
If the operators are singlets under the internal symmetry acting on $X_{1,2}$, then
they will not affect the $X_{1,2}$ geometry and only $AdS_5$ will get deformed.
 Otherwise the internal Killing symmetry will generically break
and the geometry of the spaces $X_{1,2}$ will be deformed.

There are  bulk scalar fields of the two theories corresponding to $O_{\Delta}$, $\tilde O_{4-\Delta}$
that we denote
by $\Phi_{\Delta}$ and $\tilde \Phi_{4-\Delta}$.
They have the same mass, $m^2\ell_{AdS}^2=\Delta(\Delta-4)$.
Their asymptotic behavior close to the associated boundaries
$r_{1,2}\to 0$ is\footnote{Note that we have been
 careful to separate the radial coordinates $r_1$ and $r_2$
as they belong to different $AdS_5$ spaces.}
\be
\Phi_{\Delta}\sim q_1(x)r_1^{\Delta}+p_1(x)r_1^{4-\Delta}\sp \tilde
 \Phi_{4-\Delta}\sim p_2(x)r_2^{\Delta}+q_2(x)r_2^{4-\Delta}
\label{10}\ee
where we have assumed without loss of generality that $\Delta\leq 2$.
$p_1(x)$ and $p_2(x)$ correspond to the expectation values of the associated operators
while $q_1(x)$ and $q_2(x)$ correspond to sources.

The appropriate  boundary conditions that implement the deformation of the product theory generated by (\ref{9})
are\footnote{We use the conventions of \cite{muck} that differ by a sign from those of \cite{witten}.}
\be
q_1(x)+g~p_2(x)=0\sp q_2~(x)+g ~p_1(x)=0
\label{11}\ee
where we generalized slightly the discussion in \cite{witten}.

This perturbation is exactly marginal, to leading order in $1/N_{1,2}$, \cite{witten}.
\footnote{In this case, when the two CFT's are the same, we have a $Z_2$ exchange
symmetry. This provides  sufficient conditions, for the $g\to 1/g$ duality advocated in \cite{witten}
to hold (g is the coupling parameterizing the marginal line).}
The ``background" solution consistent with the boundary conditions (\ref{11}) is the trivial solution:
 \be
\Phi_{\Delta}(x,r_1)=\tilde\Phi_{4-\Delta}(x,r_2)=0\;\;.
\label{16}\ee
Therefore the background remains unchanged. The sole modification emerges from the coupled boundary conditions,
 and we will investigate below how we may compute correlation function in the bulk theory.

It is  obvious that the two bulk geometries must remain distinct,
since a priori the original CFTs have different bulk geometries.
They share however the same four-dimensional coordinates
at their boundary, since these correspond to the common coordinates of the two CFTs.
 It is convenient to think of the two geometries as ``glued" at their boundary, via
the correlated boundary conditions (\ref{11}). This is however not a geometric junction in the usual sense.
It should be thought of as  a common surface were correlated boundary conditions are imposed.
However, the common boundary is a traversable surface for light signals. Indeed a light signal travelling towards the boundary of one of
the AdS spaces reaches the boundary in finite time. The correlated boundary conditions allow the signal to cross into the other AdS space
and continue its travel towards the center.

The background solution for the perturbing scalars, breaks the conservation of the individual stress tensors of each CFT.
Only the sum is guaranteed to be conserved by overall translation invariance, and therefore only the graviton coupling to the overall
stress tensor is massless. The orthogonal linear combination becomes massive as advocated earlier.
This state of affairs can be favorably compared to the analysis of a graviton mass in AdS in \cite{porrati}.
The mass-generating boundary conditions for the scalars, are essentially half of the story we have here.
The perturbation conditions (\ref{11}) reproduce the transparent conditions in \cite{porrati} in one of the CFTs
responsible for giving a mass to the graviton. Here, having two copies of the scalars,
we end up with one massive and one massless graviton instead.

The general solution of the (linear)  equations for $\Phi_\Delta$ and $\Phi_{4-\Delta}$,
subject to the boundary conditions (\ref{11}) Fourier-transformed in the four-dimensions can be written as
\be
\Phi_{\Delta}(\vec p,r_1)=r_1^2\left[a(\vec p)I_{\nu}(|\vec p|r_1)+b(\vec p)K_{\nu}(|\vec p|r_1)\right]
\sp \nu=\sqrt{4+m^2\ell_{AdS}^2}=|\Delta-2|
\label{399}\ee
\be
\tilde \Phi_{4-\Delta}(\vec p,r_2)=r_2^2\left[ {a(\vec p)\over g}I_{\nu}(|\vec p|r_2)+gb(\vec p)K_{\nu}(|\vec p|r_2)\right]
\label{409}\ee
They correspond to boundary conditions in \cite{porrati} $\a=g,\b=1/g$.

The Witten-Gubser-Klebanov-Polyakov prescription for calculating correlators in the perturbed theory goes through
with some  modifications, \cite{witten}-\cite{minces}.
We will describe this first in the known case of multitrace deformations inside a single CFT.
We denote the normalized generating singe-trace operator by $O(x)$, of dimension $\Delta<2$, and the dual bulk scalar field by $\Phi$.
The perturbed CFT action is
\be
I^W=I_{CFT}+\int d^4 x ~W(O)
\label{18}\ee
where $W(O)$ is a local functional, that is linear in the case of single trace perturbations, but non-linear for the
multitrace ones.
The CFT action is related to the bulk supergravity action as
\be
\langle \exp\left[-\int d^4x ~\a ~O\right]\rangle =\exp\left[-I_{sugra}(q)\right]
\label{19}\ee
where the source $\a(x)$ is related to the asymptotic form of the bulk field $\Phi$ as
\be
\lim_{r\to 0}\Phi(x,r)\sim r^{\Delta}q(x)+r^{4-\Delta}p(x)+\cdots\sp q(x)+\a(x)=0
\label{20}\ee
The bulk action is naturally a functional of $q$.
In the Hamilton-Jacobi formalism, $p$ and $q$ are conjugate variables with
\be
p=-{\delta I_{sugra}(q)\over \delta q}\sp q={\delta J(p)\over \delta p}\sp J(p)=I_{sugra}-\int d^4x~qp
\label{21}\ee
The appropriate bulk generating functional for the perturbed theory is
\be
I^W_{sugra}(\a)=I_{sugra}(q)+\int d^4x~\left (W(p)-p{\delta W\over \delta p}\right)
\label{22}\ee
with $p,q$ related to the source $\a$ by (\ref{21}) and
\be
{\delta I^W_{sugra}\over \delta p}=q+{\delta W(p)\over \delta p}+\a=0
\label{23}\ee
Therefore the bulk/boundary correspondence translates to
\be
\langle \exp\left[-\int d^4x~\a~O\right]\rangle_W=\exp\left[-I^W_{sugra}(\a)+I^W_{sugra}(0)\right]
\label{24}\ee
where the boundary theory expectation value is taken in the W-deformed CFT.

We now consider the case of interest to us, namely, two CFTs interacting via (\ref{9}).
The perturbed CFT action is
\be
I^W=I_{CFT_1}+I_{CFT_2}+\int d^4 x ~W(O_{\Delta},\tilde O_{4-\Delta})\sp
 W(O_{\Delta},\tilde O_{4-\Delta})=g~O_{\Delta}\tilde O_{4-\Delta}\eqp
\label{25}\ee
In the Hamilton-Jacobi formalism, we now have two independent variables $p_i$ and $q_i$
defined in (\ref{10}) with
\be
p_i=-{\delta I^i_{sugra}(q_i)\over \delta q_i}\sp q_i={\delta J^i(p_i)\over \delta p_i}\sp J^i(p)=I^i_{sugra}-\int d^4x~q_ip_i\sp i=1,2
\label{26}\ee
and $i$ is not summed. $I^{1,2}_{sugra}(q_{1,2})$ are the two supergravity actions of the associated decoupled CFTs.
The appropriate bulk generating functional for the perturbed theory is now
\be
I^W_{sugra}(\a_1,\a_2)=I^1_{sugra}(q_1)+I^2_{sugra}(q_2)+\int d^4x~\left (W(p_1,p_2)-\sum_{i=1}^2p_i{\delta W\over \delta p_i}\right)
\label{27}\ee
with $p_i,q_i$ are determined by the  sources $\a_i$ by (\ref{26}) and
\be
{\delta I^W_{sugra}\over \delta p_i}=q_i+{\delta W(p_1,p_2)\over \delta p_i}+\a_i=q_i+g~(\s^1)^{ij}p_j+\a_i=0
\label{28}\ee
where $(\s^1)^{ij}$ is the standard Pauli matrix.
For vanishing sources $\a_i$, they reproduce (\ref{11})

The bulk/boundary correspondence recipe is
\be
\langle \exp\left[-\int d^4x~\left(\a_1~O_{\Delta}+\a_2\tilde
 O_{4-\Delta}\right)\right]\rangle_W=\exp\left[-I^W_{sugra}(\a_1,\a_2)+I^W_{sugra}(0,0)\right]
\label{29}\ee
This generalizes in a straightforward fashion to more complicated interactions.

We may also include sources $A_a,B_a$ for any other single trace operators of the two theories. These will appear in
(\ref{27}) via $I^1_{sugra}(q_1,A_a)$ and $I^2_{sugra}(q_1,B_a)$, while the interaction $W$ is independent on them.
It is straightforward to verify, by expanding in powers of the interaction, that (\ref{29})
matches the perturbative expansion of the field theory correlators.

\section{Stress tensors and gravitons}

We may now investigate the corrections to the correlators of the stress tensors and the associated interactions
of the two gravitons. We denote by $T^1_{\m\n}$ the stress-tensor of CFT$_1$, and by $T^2_{\m\n}$ the one of CFT$_2$.
We also set $N_1=N\to \infty$, and $N_2={\tt x} N$ with ${\tt x}$ finite. We will mostly neglect the ${\tt x}$ dependence in the following.

We may directly verify that the corrections to the various two point functions, are subleading in $1/N$. Schematically, the first
non-trivial corrections (after normal ordering the interaction to second order)
occur at order ${\cal O}(g^2)$,\footnote{There are also additive renormalizations
due to corrections at the one-point function. They are of the same order, and we subtract them from the associated operators.}
\be
\delta \langle T^1(x)T^1(y)\rangle={g^2\over 2!}~\int d^4z_1d^4 z_2~\langle
T^1(x)T^1(y)O(z_1)O(z_2)\rangle_c~\langle \tilde O(z_1)\tilde O(z_2)\rangle
\label{b30}\ee
\be
\delta \langle T^1(x)T^2(y)\rangle={g^2\over 2!}~\int d^4z_1d^4 z_2~\langle
T^1(x)O(z_1)O(z_2)\rangle_c~\langle T^2(y)\tilde O(z_1)\tilde O(z_2)\rangle
\label{b31}\ee
\be
\delta \langle T^2(x)T^2(y)\rangle={g^2\over 2!}~\int d^4z_1d^4 z_2~\langle
T^2(x)T^2(y)\tilde O(z_1)\tilde O(z_2)\rangle_c~\langle  O(z_1) O(z_2)\rangle
\label{b32}\ee
These corrections are of order ${\cal O}\left({g^2\over N^{2}}\right)$ and therefore the graviton mass is of the same order.
Note also that the correction in (\ref{b31}) is a trivial ($x,y$-independent) wave-function
renormalization.
We conclude  that the corrections to the two-point functions of the two original stress tensors are subleading in $1/N$.
This is not the case however for higher-point functions.
Consider the deformation of the three-point functions. The following
$$
\delta \langle T^1(x_1)T^1(x_2)T^1(x_3)\rangle={g^2\over 2!}~\langle T^1(x_1)T^1(x_2)\rangle
\times
$$
\be
\times\int d^4z_1d^4 z_2~\langle
T^1(x_3)O(z_1)O(z_2)\rangle_c~\langle \tilde O(z_1)\tilde O(z_2)\rangle
\label{b33}\ee

$$
\delta \langle T^1(x_1)T^1(x_2)T^2(x_3)\rangle={g^2\over 2!}~\langle T^1(x_1)T^1(x_2)\rangle
\times
$$
\be
\times\int d^4z_1d^4 z_2~\langle
T^2(x_3)\tilde O(z_1)\tilde O(z_2)\rangle_c~\langle  O(z_1) O(z_2)\rangle
\label{b34}\ee
is a correction that is removed by the standard shift of the stress tensors.
The connected contribution is
$$
\delta \langle T^1(x_1)T^1(x_2)T^1(x_3)\rangle=
$$
\be
={g^2\over 2!}~\int d^4z_1d^4 z_2~\langle T^1(x_1)T^1(x_2)
T^1(x_3)O(z_1)O(z_2)\rangle_c~\langle \tilde O(z_1)\tilde O(z_2)\rangle
\label{b36}\ee

$$
\delta \langle T^1(x_1)T^1(x_2)T^2(x_3)\rangle=
$$
\be
={g^2\over 2!}~\int d^4z_1d^4 z_2~\langle
T^2(x_3)\tilde O(z_1)\tilde O(z_2)\rangle_c~\langle  T^1(x_1)T^1(x_2)O(z_1) O(z_2)\rangle_c
\label{b35}\ee

This correction is of order ${\cal O}\left({g^2\over N^3}\right)$. The unperturbed result is
${\cal O}\left({1\over N}\right)$.
More generally, the leading correction to the correlation function of n stress tensors,
 is given by a factor of ${g^2\over N^2}$ multiplying the unperturbed result.

We therefore expect that in the bulk theory, at leading order, the propagators of the two graviton $h_{\m\n}^{1,2}$  are unchanged,
while there are non-trivial changes in their interactions.
To next order, the graviton propagators are modified, $h^1+h^2$ remains massless, while $h^1-h^2$ acquires  a mass,
according to our previous arguments.
This is in agreement with \cite{porrati} where it was found that
\be
m_{\rm graviton}^2\sim {1\over N^2\ell_{AdS}^2}.
\label{411}\ee
Using the AdS/CFT dictionary
\be
M^3\sim {N^2\over \ell_{AdS}^3}
\label{421}\ee
where $M$ is the five-dimensional Planck scale,
 we do indeed observe that the graviton mass
is suppressed by a factor $1/N^2$ compared to the kinetic term.

In the sequel, we will investigate specific examples of the general picture painted above.

\section{Examples in four dimensions\label{4d}}

\subsection{Coupling two ${\cal N}=4$ super Yang Mills theories}

We first consider the  case where the two CFTs coupled non-trivially via
an interaction that is relevant in the UV\footnote{The opposite case, namely an irrelevant perturbation,
 is easier to come by, and is more or less trivial for our purposes:
for example turning on scalar expectation values is ${\cal N}=4$ sYM, we may break in the IR the
gauge group to $U(N_1)\times U(N_2)$ with $N_{1,2}\gg 1$. However, it leads to the standard  multi-throat
geometries in the IR with very interesting associated physics.} are both  ${\cal N}=4$ sYM.

The gauge invariant normalized operators with the minimum free-field dimensions are
\be
O=\sum_{I=1}^6 Tr[\Phi^I\Phi^I]\sp O_{IJ}\equiv
\left[Tr[\Phi^I\Phi^J]-{1\over 6}\delta^{IJ}~O\right]
\label{6}\ee
The Konishi operator $O$, is non-BPS and is therefore known to have a large anomalous
in the strong coupling limit $\l\to\infty$.
$O_{IJ}$ are BPS operators and their scaling dimension remains $\Delta=2$,  as they are protected.
We must construct an interaction between two copies $CFT_{1,2}$ of ${\cal N}=4$ sYM
with 't Hooft couplings, $\l_{1,2}$ and number of colors $N_{1,2}$ not necessarily equal.
The only interaction that might be marginal or marginally-relevant is
\be
S_{\rm interaction}=h_{IJ,KL}\int d^4x~O_{IJ}\tilde O_{KL}
\label{7}\ee
where $O_{IJ}\in CFT_1$ and $\tilde O_{KL}\in CFT_2$.
Classically  this interaction is marginal. According to our discussion in appendix \ref{A},
it is marginally relevant at one loop.

Unfortunately, the theory with the interaction in (\ref{7}) is non-perturbatively unstable.
(\ref{7}) is an addition to the potential of the scalars of the two CFTs.
We will consider a configuration of the scalars of the two theories so that only fields in the Cartan are non-zero.
We denote their eigenvalues  as $\Phi^I_i$, $i=1,2,\cdots,N_1$, $\tilde\Phi^I_i$, $i=1,2,\cdots,N_2$.
For such scalar values, the potentials of the two CFTs vanish.
The interaction (\ref{7}) now becomes
\be
S_{\rm interaction}={h_{IJ,KL}\over N_1N_2}\int d^4x~
\left[\Phi^I\cdot \Phi^J-{1\over 6}\delta^{IJ}\Phi\cdot \Phi\right]
\left[\tilde\Phi^I\cdot \tilde\Phi^J-{1\over 6}\delta^{IJ}\tilde\Phi\cdot \tilde\Phi\right]
\label{8}\ee

It is easy now to see that both diagonal and off-diagonal elements in $O_{IJ}$ can be made positive or negative
and arbitrarily large
by appropriate choices of the Cartan values. Therefore, the potential in (\ref{8})
 has directions where it becomes arbitrarily negative or positive, for all couplings $h_{IJ,KL}$.
Therefore this perturbation destabilizes the theory.
 We conclude that we can couple two ${\cal N}=4$ super Yang Mills theories via a marginally relevant perturbation in the UV,
 however this coupling is not well defined non-perturbatively.

\subsection{Coupling two  ${\cal N}$=1 $T^{1,1}$ conifold theories}

This CFT involves the quiver ${\cal N}=1$ SU(N)$\times $SU(N) gauge theory with
two bifundamental chiral multiplets $A_{i}$, $i=1,2$, and two anti-bifundamental
chiral multiplets $B_i$, \cite{kw}.
This theory is expected to flow in the IR to a strongly-coupled fixed-point theory
with its global symmetry SU(2)$\times $SU(2)$\times$U(1)$_R$ intact.
It is a line of fixed points parameterized by a combination of the two coupling constants.
Its dual gravitational theory is described by the geometry $AdS_5\times T^{1,1}$
where the five-dimensional manifold $T^{1,1}$ is the coset $(SU(2)\times SU(2))/U(1)$
realizing the global symmetry of the theory as its isometry.

The theory can be deformed by an SU(2)$\times $SU(2)$\times$U(1)$_R$-invariant  superpotential
\be
W\sim Tr[A_1B_1A_2B_2-A_1B_2A_2B_1]
\label{14}\ee
while keeping the conformal symmetry , after some adjustment of the rest of the couplings.
This provides a two parameter family of SU(2)$\times $SU(2)$\times$U(1)$_R$-invariant  CFTs, \cite{kw}.

In \cite{nonlocal} it was observed that the double-trace deformation generated by the superpotential
\be
\tilde W\sim Tr[A_1B_1]Tr[A_2B_2]-Tr[A_1B_2]Tr[A_2B_1]
\label{15}\ee
is exactly marginal, and introduced a new parameter in the CFT without breaking the global symmetry.
We will use this observation to generate a marginal deformation coupling two copies of the conifold
quiver theory.

 Consider the product of two copies of the conifold theory, $CFT\times \tilde{CFT}$ and a superpotential
\be
\hat W\sim Tr[A_1B_1]Tr[\tilde A_2\tilde B_2]-Tr[A_1B_2]Tr[\tilde A_2\tilde B_1]
\label{166}\ee
If the two CFTs are at the same point in the moduli space, then the arguments of \cite{nonlocal} imply the marginality
of this perturbation.
This breaks the (SU(2)$\times $SU(2)$\times$U(1)$_R)^2$ symmetry of the decoupled theories to the diagonal one.
In particular this implies that the associated gauge group (SU(2)$\times $SU(2)$\times$U(1)$_R)^2$ of the
product theory will be Higgsed to the diagonal one. This is similar to the fate of the two gravitons.

The scalars $\Phi_{ij}$ and $\tilde \Phi_{ij}$, relevant for the deformation transform in the (2,2)$_0$
representation of the (SU(2)$\times$SU(2)$\times$U(1)$_R$ global symmetry.
They are trivial in the background solution, and the geometry therefore remains (AdS$_5\times T^{1,1})^2$.
The symmetry however is broken to the diagonal one by the boundary conditions that break the gauge symmetry
giving a mass to half of the bulk gauge bosons.

\section{Examples in two dimensions}

There are many  examples in two dimensions, realizing the general framework exposed above.
Moreover, in two dimensions, there are non-trivial couplings between two distinct large-N CFTs
generated by perturbing operators that are products of currents.

To be concrete consider a large-N CFT$_1$ which has a conserved chiral U(1) current, $J_1$.
Consider also another large-N CFT$_2$ with an anti-holomorphic U(1) current $\bar J_2$.
We may now consider the perturbation
\be
\delta S=g\int d^2z~J_1\bar J_2\label{60}
\ee
that couples the two CFTs with the
 well-known effect of providing an O(1,1) boost on the associated charge lattice.
The question whether the current operators are single-trace is
 irrelevant here as all connected higher-point correlation
 functions
of U(1) currents vanish.

There are also examples that involve couplings with scalar operators.
In appendix \ref{2d} we have analyzed a large-N two-dimensional
conformal gauge theory that is isomorphic to the conformal coset
CFT
\be
SU(N)_{k_1}\times SU(N)_{k_2}\over SU(N)_{k_1+k_2}
\label{61}\ee
This is obtained by gauging the diagonal SU(N) symmetry of the
CFT $SU(N)_{k_1}\times SU(N)_{k_2}$ and adding a kinetic term
for the SU(N) gauge fields. In the IR, the coupling of this term
flows to zero and the IR theory is the conformal coset
in (\ref{61}).
The relevant parameters here are the number of colors N and the 't Hooft coupling constants
\be
\l_1={N\over k_1}\sp \l_2={N\over k_2}
\label{62}\ee
The large-N limit is $N\to \infty$ with $\l_i$ kept fixed.
Most primary operators are in one-to-one correspondence with
triplets\footnote{There are exceptions to this rule, but they will not be relevant here.}
 of representations $(R_1,R_2,R)$
with $R_1\in SU(N)_{k_1}$, $R_2\in SU(N)_{k_2}$, $ R_1\otimes R_2\sim R\in SU(N)_{k_1+k_2}$.
Their holomorphic conformal dimension is
\be
\Delta_{R_1,R_2;R}={C_2(R_1)\over k_1+N}+{C_2(R_2)\over k_2+N}-{C_2(R)\over k_1+k_2+N}
\label{63}
\ee
For the diagonal modular invariant they correspond to scalar
operators with scaling dimension, twice that of (\ref{62}).
In appendix \ref{2d} we show that operators associated to ($R,\bar R,X$)
 with $X\in R\otimes \bar R$ are single trace operators.
Consider therefore the product of this CFT, with $N_1$ colors and $\l_1=\l_2=\l$ and
CFT' with $N_2$ colors and $\l_1'=\l_2'=\l'$
and the class of single-trace operators with
\be
\Delta_{R_k,\bar R_k;1}=k{\l\over \l+1}+{\cal O}\left({1\over N}\right)
\label{64}\ee
where $R_k$ is the $k$-index symmetric tensor of SU(N).
Then the perturbation
\be
\delta S=g\int d^2z ~ \Phi_{R_k,\bar R_k;1}\Phi'_{R_l,\bar R_l;1}
\label{65}\ee
is marginally relevant if we choose
\be
\l'={1+(1-k)\l\over (l-1)+(l+k-1)\l}\sp k>1 ~~{\rm or}~~ l>1
\label{66}\ee
There are other straightforward possibilities of large-N nearly
marginal couplings but we will not pursue them further here.

\section{Multiply coupled CFTs}

We have seen so far that we can couple two large-N CFTs with marginal perturbations giving rise in the dual
description to coupled string theories on a product space, with two gravitons.

It is straightforward to extend this to coupling of more than two large-N CFTs.
Consider $M$ such CFTs, CFT$_i$, $i=1,2,\cdots, M$.
For each pair of CFTs, with one containing a single-trace operator $O^i_{\Delta}$ of dimension $\Delta$ and the other
$O^j_{D-\Delta}$ of dimension $D-\Delta$
we may write a coupling via a perturbation $g_{ij}\int ~O^i_{\Delta}O^j_{D-\Delta}$.
Therefore,
\be
W=\sum_{<ij>}g_{ij}p_ip_j
\label{35}\ee
where the sum extends to all pair where conjugate operators exist.
This defines a graph, where the nodes are the CFT's, a link between two nodes
indicates the existence of such conjugate perturbations.
A multiple link indicates the presence of more than one such couplings.
The Hamilton-Jacobi formalism exposed in section \ref{inter} generalizes in a straightforward fashion
to this case.

An interesting question is whether more than two CFT's can be coupled together simultaneously.
The answer to this question depends crucially on the space-time dimension.
Consider the product of several CFTs in four dimensions: CFT$_i$, and a relevant perturbation of the form
\be
\delta S=g\int d^4x ~\prod_{i=1}^n ~\Phi_{\Delta _i}
\label{36}\ee
with $\sum_{i=1}^n\Delta_i\leq 4$. The unitarity bound in four dimensions, $\Delta_i\geq 1$, implies that at most
four such operators can be used. However in the maximal  case all must have $\Delta=1$ and
 the group theory requires that they are free scalars.
The perturbation (\ref{36}) is then an unstable potential and such a deformation is not well defined beyond perturbation theory.
The only remaining case is three CFT's with operators of dimension $1<\Delta <{4\over 3}$.

There are four-dimensional CFTs that contain such operators.
Consider, SQCD, in the conformal window ${3\over 2}<{N\over N_f}\leq 3$ and its associated IR CFT, \cite{seiberg}.
In this CFT the meson operators have scaling dimension $\Delta_m =3-3{N\over N_f}$, and therefore
satisfy $1\leq \Delta_m\leq 2$.
In this case we must take the large N limit by also scaling the number of flavors:
\be
N\to \infty\sp N_f\to \infty\sp  {N\over N_f}\to {\tt x}~~~{\rm fixed}
\label{37}\ee
The dual geometry was argued to be $AdS_5\times S^1$, realized in non-critical string theory, \cite{KM}.
If we consider the product of three such CFTs: CFT$_{N_i,{\tt x}_i}$, with ${\tt x_1+x_2+x_3}=1$, we may write a marginal
perturbation that couples all three theories together
\be
\delta S=g\int d^4x ~\Phi_{\Delta_{m_1}}\Phi_{\Delta_{m_2}}\Phi_{\Delta_{m_3}}
\label{38}\ee

The discussion of section \ref{inter} carries over here as well, and the Hamilton-Jacobi formalism
generalizes straightforwardly.
The geometrical interpretation is a bit more exotic.
We will have to consider the product of three $AdS_5\times S^1$ geometries coupled via their common $S^4$ boundary
via the appropriate boundary conditions.
The rough picture is that of a 3-junction, with the common point associated with the common boundary $S^4$ and the three
legs associated with the three $AdS_5$ holographic coordinates as well as the three circles.

Therefore, for four-dimensional boundaries, only two- and three-junctions exist.

In other dimensions the situation can be very different.
In six dimensions the unitarity bound is $\Delta\geq {D-2\over 2}=2$.
Therefore, a three-junction exists only for operators at the lower bound, with $\Delta=2$. Again these are free scalars, and the perturbation
is that of an unstable cubic interaction. We conclude that higher than binary junctions do not exist non-perturbatively in six-dimensional CFTs.

In two dimensions  the unitarity bound on operators is $\Delta>0$, and in principle, operators with very low dimension
exist. This suggests that we may have arbitrary k-junctions of CFT's in two dimensions.
Taking the example of the large-N CFT described in detail in appendix \ref{2d}, the single trace operators
$\Phi_{\Yboxdim4pt\yng(1),\overline{\Yboxdim4pt\yng(1)};1}$ have large-N scaling dimension
\be
\Delta_{\Yboxdim4pt\yng(1),\overline{\Yboxdim4pt\yng(1)};1}=
{1\over 2}\left[{\l_1\over 1+\l_1}+{\l_2\over 1+\l_2}\right]
\ee
To construct an arbitrary k-junction, with $k>1$, we may choose the two couplings of each of the k-copies of the CFT
as $\l_1=\l_2={1\over k-1}$.
Then, the operator in question, has scaling dimension $2\Delta_{\Yboxdim4pt\yng(1),\overline{\Yboxdim4pt\yng(1)};1}={2\over k}$.
Therefore a product of k of them, one from each CFT is marginal at large N.

\section{Outlook}

We have analyzed the bulk/boundary correspondence in cases where multiple gravitons are present in the bulk theory.
This arises when the CFT in the UV is a product of two or more large-N CFTs.
In the absence of cross interactions the bulk theory is described by the direct sum of two gravitational (string) theories.

The CFTs can be coupled together by a double-trace marginal or relevant perturbation.
The effect of such a coupling is to preserve a massless graviton,
but render all other gravitons massive.
The graviton mass is subleading compared to its kinetic term, as it originates from non-planar diagrams.
On the other hand, interactions of the massive graviton are modified to leading order.

Such bulk backgrounds are described by products of AdS spaces times non-compact manifolds.
The AdS spaces are identified at their boundaries.
Moreover, there are coupled boundary conditions for the various operators at the common boundary.

In this context one can also answer the question that has been asked in several previous contexts: are there
theories of two or more massless interacting gravitons. Several arguments from field theory and
string theory indicate that the answer to this question is no.
We argue, using the bulk-boundary correspondence that the answer here is also negative.
However, several massive gravitons interacting with a massless one are
 possible, as shown by direct construction.

An interesting aspect of the above is the picture that emerges for the set of all geometries dual to large-N QFT.
It is obvious that the space of such geometries has a structure similar to cobordism,
although this analogy may be misleading in the details.
In particular we can ``glue" two spaces by identifying their common boundary,
if there are appropriate matching operators in the two associated CFTs.
We also  assign concrete coupled boundary conditions for the associated bulk theories.
Moreover here we have also the concept of three- and higher junctions (relevant for two dimensional boundaries)
arising from interacting higher products of CFTs.
It is a very interesting project to analyze this structure further.

We should also stress that this picture, is genuinely different from the case of multi-throat single manifolds, popular in
string compactifications. Such manifolds, contain a single metric, and arise from the split of the large-N theory in the IR.
In such cases, it has been argued \cite{dimo} that tunneling between such throats may provide small numbers.
It is interesting to investigate this question in our context.

We have seen in the example of ${\cal N}=4$ sYM theories in four dimensions that perturbative couplings of two large-N
CFTs may be non-perturbatively unstable. It is an interesting open question whether this is a generic phenomenon.

Another interesting question concerns bulk black holes and finite temperature effects in the boundary theory.
On the field theory side we may put the two non-interacting CFTs in a thermal state, corresponding to different a priori temperatures
$T_1\not =T_2$. Once we couple the two CFTs via the double-trace deformation, this will give rise to a metastable state
that will eventually relax to a common temperature $T$. In the product space, the original configuration would correspond
to two independent AdS black holes, if $T_{1,2}>T_{\rm deconf}$. If one of the temperatures is below the deconfinement transition, then we must substitute
the AdS black-hole by thermal AdS.
Once the two CFTs are coupled, the common boundary (now with topology $S^1\times S^3$) allows only static configurations with
a common inverse temperature, $\beta$, equal to the radius of the common boundary $S^1$.
It is an interesting question whether the equilibration of an initial state with different temperatures can be achieved via bulk physics exchanging
energy through the common boundary. Another interesting question is whether the idea of massive spin-two black hole hair, \cite{dvali}
can be implemented in this context.

The geometric picture of distinct interacting string theories in
asymptotically AdS spaces, can also be entertained
in the context of standard asymptotically flat geometries.
We may consider two district string theories, which share the same asymptotic
infinity, but their "interiors" are distinct.
Consider a string theory $ST_1$  in a vacuum of the form $M_4\times C_6$ where $M_4$ is Minkowski space
with the usual asymptotic boundary $\partial M_4$.
We also consider a second string theory $ST_2$ in a vacuum of the form
$\tilde M_4\times \tilde C_6$ where $M_4$ is Minkowski space.
The two string theories can have different parameters, $g_{s}\not= g_s'$ and $\ell_s\not=\ell_s'$.
They can be coupled by the product of two
massless perturbations (the analogue of marginal perturbations in the
AdS case).
This will correlate the  scattering amplitudes in the two theories in a  fashion similar
to the AdS case.

The interpretation of the space-time physics in such a context and its implications for the large scale structure of
the observable universe remain to be understood.

\vskip 2cm
\begin{flushleft}
{\large \bf Note added}
\end{flushleft}

I am aware that O. Aharony, A. Clark and A. Karch have been pursuing similar ideas, see \cite{ack}.

\noindent

\vskip 2cm
\begin{flushleft}
{\large \bf Acknowledgments}
\end{flushleft}

\noindent

I would like to thank T. Petkou, who
participated in early stages of this paper, for discussions.
Many thanks to  O. Aharony,  M. Bianchi, S. Dimopoulos, K. Intriligator, I. Klebanov and W. M\"uck for discussions.
I would also like to thank the Galileo Galilei Institute for hospitality during the last stages of this work.
The work was partially supported by ANR grant NT05-1-41861,
INTAS grant, 03-51-6346,
RTN contracts MRTN-CT-2004-005104 and MRTN-CT-2004-503369,
CNRS PICS 2530 and 3059 and by a European Excellence Grant,
MEXT-CT-2003-509661.

 \newpage
\appendix

 \renewcommand{\theequation}{\thesection.\arabic{equation}}
\addcontentsline{toc}{section}{Appendices}
\section*{APPENDIX}

\section{CFT perturbations to leading order in ${1\over N}$\label{A}}

In this appendix we detail the structure of CFT deformations by  single and double trace  operators.
Let $O_{\Delta}\sim Tr[\cdots]$ be a single trace operator of dimension $\Delta$, normalized
so that its two-point function is
\be
\langle O_{\Delta}(x)O_{\Delta}(y)\rangle={1\over |x-y|^{2\Delta}}
\label{a1}\ee
It is well known that for single trace operators, the connected higher-point functions are suppressed
at large $N$,
\be
\langle \prod_{i=1}^n ~O_{\Delta_i}(x_i)\rangle_c \sim N^{2-n}
\label{a2}\ee
It is also known that single-trace operators can also mix with multiple-trace  operators.
For our purposes this can be neglected as it is subleading  in ${1\over N}$ generically.
It is only in the case of degenerate dimensions that the mixing can be of order one.

Consider now a perturbation of the CFT by
\be
\delta S=g\int d^d x~ O_{\Delta}(x)
\label{a3}\ee
with $\Delta \leq d$.
Assume first that $g\sim {\cal O}(1)$.
We will also assume that the dimension $\Delta$ is known exactly. This may happen because
the operator is BPS and therefore protected. In two dimensions, this is not necessary,
as we have control over a larger range of CFTs.
Because of (\ref{a2}), to leading order in ${1\over N}$,
the perturbing operator has only disconnected n-point functions.
Therefore, upon renormalization, its dimension remains constant as we vary the coupling $g$.
To see this we evaluate the n-th order correction to its two-point function as
\be
g^n\langle O_{\Delta}(x)O_{\Delta}(y)\prod_{i=1}^n ~\int d^d z_i~ O_{\Delta_i}(z_i)\rangle=
Z_1\langle O_{\Delta}(x)O_{\Delta}(y)\rangle+Z_2
\label{a4}\ee
\be
Z_1=g^n \langle \prod_{i=1}^n ~\int d^d z_i~ O_{\Delta_i}(z_i)\rangle
\label{a6}\ee
\be
 Z_2=n(n-1)g^n\left[ \langle O_{\Delta}(x)\int d^d y O_{\Delta}(y)\rangle\right]^2
\langle \prod_{i=1}^{n-2} ~\int d^d z_i~ O_{\Delta_i}(z_i)\rangle
\label{a5}\ee
$Z_{1,2}$ are in general cutoff dependent constants .
By redefining the perturbing operator order by order in perturbation theory as
\be
O_{R}=\sqrt{Z_1}~O_{\Delta} (x)+\sqrt{Z_2}
\label{a7}\ee
we deduce that its scaling dimension remains intact.
An alternative way to renormalize, is to normal order the exponential of the
interaction, $:e^{g\int d^d x~ O_{\Delta}(x)}:$.

However, the same argument indicates that no other operator changes dimension, or mixes.
Therefore this perturbation is trivial to leading order.

We now examine the more interesting case where $g=hN\sim {\cal O}(N)$.
To zero-th order there is a linear contribution to the $\beta$-function if $\Delta\not =d$
\be
{\pa h\over \pa \log \mu}\equiv\beta(h)=(d-\Delta) h+\cdots
\label{a8}\ee
If $\Delta=d$, the classical $\beta$-function is zero and the perturbation marginal.
There can be however a next-to-leading  (one-loop) contribution to the two-point function of the perturbing operator
\be
\delta\langle O_{d}(x)O_{d}(y)\rangle=hN\langle O_{d}(x)O_{d}(y)~\int d^d z~O_{d}(z)\rangle={h\over |x-y|^{d}}\int d^d z~
{C_{ddd}\over (|x-z||y-z|)^{d}}
\label{a9}\ee
that may contribute to the $\beta$-function to next order. If
 $C_{ddd}$ is non-zero it is typically of order ${\cal O}(1)$.
Inserting a UV cutoff $a$ we obtain for the logarithmic divergence
\be
\delta\langle O_{d}(x)O_{d}(y)\rangle={2\pi^{d\over 2}hC_{ddd}~\over \Gamma\left({d\over 2}\right)}
{1\over |x-y|^{2d}}\log{|x-y|^2\over a^2}+\cdots
\label{a999}\ee
which implies an anomalous dimension for the perturbing operator
\be
\Delta_{R}=d-{\pi^{d\over 2}\over \Gamma\left({d\over 2}\right)}hC_{ddd}+{\cal O}(h^2)
\label{a10}\ee
and an associated $\beta$-function
\be
\beta(h)=-{\pi^{d\over 2}\over \Gamma\left({d\over 2}\right)}~C_{ddd}~h^2+{\cal O}(h^3)
\label{a11}\ee
The perturbation is relevant if $C_{ddd}>0$.

Consider now a perturbation by a double-trace operator $\Phi\equiv O_{\Delta_1}O_{\Delta_2}$
where $O_{\Delta_i}$ are single trace operators,
\be
\delta S=g\int d^d x~ :~O_{\Delta_1}(x)O_{\Delta_2}(x):
\label{a12}\ee
We will set $\Delta_2=d-\Delta_1\equiv d-\Delta$ as this is the case  of direct interest in this paper.
Similar estimates as above indicate that the effects of this perturbation can be non-trivial if
$g\sim {\cal O}(1)$. This scaling guarantees that the free energy scales as $N^2$.
The first non-trivial correction to the two-point function of the perturbing operator is
\be
\delta\langle \Phi(x)\Phi(y)\rangle=g\langle O_{\Delta}(x)O_{\Delta}(y)\rangle ~\int d^d z~
\langle O_{4-\Delta}(y)O_{4-\Delta}(z)\rangle \langle O_{4-\Delta}(x)O_{\Delta}(z)\rangle
+{\cal O}(N^{-2})
\label{a14}\ee
For the case $\Delta\not= d/2$, the leading contribution vanishes as $\langle O_{\Delta}O_{d-\Delta}\rangle=0$.
We therefore obtain a line of fixed points, valid to leading order in 1/N, \cite{witten}.
When $\Delta= d/2$, then from (\ref{a14}),
\be
\delta\langle \Phi(x)\Phi(y)\rangle=
{2\pi^{d\over 2}~g~\over \Gamma\left({d\over 2}\right)}
{1\over |x-y|^{2d}}\log{|x-y|^2\over a^2}+\cdots
\label{a9991}\ee
and we obtain the same $\beta$-function as in (\ref{a11}) with $C_{ddd}=1$.
The perturbation is relevant and the theory asymptotically free, \cite{witten}.

We now proceed to discuss double-trace perturbations, that couple two conformal field theories, $CFT_1$ and $CFT_2$.
Consider first properly normalized operators, of exact dimensions $\Delta_i=\Delta_I=d/2$,
\be
\langle O_i(x) O_j(y)\rangle={\delta_{ij}\over |x-y|^{d}}\sp \langle O_I(x) O_I(y)\rangle={\delta_{IJ}\over |x-y|^{d}}
\ee
where $O_i\in CFT_1$, $O_I\in CFT_2$.
Consider now the perturbing double trace operators, $\Phi_{ij}=:O_iO_j:$, $\tilde\Phi_{IJ}=:O_IO_J:$, $X_{iI}=:O_iO_I:$.
We obtain the following structure constants at leading order in 1/N from the connected three-point function\footnote{Note that
in terms of the individual single-trace operators $O_i$, $O_I$, this is the disconnected component. The truly connected component
is subleading in 1/N.}
\be
C_{\Phi_{ij}\Phi_{kl}\Phi_{mn}}=\delta_{ik}(\delta_{jm}\delta_{ln}+\delta_{jn}\delta_{lm})
+\delta_{il}(\delta_{jm}\delta_{kn}+\delta_{jn}\delta_{km})+
\ee
$$
+\delta_{im}(\delta_{jk}\delta_{ln}+\delta_{jl}\delta_{kn})
+\delta_{in}(\delta_{jk}\delta_{lm}+\delta_{jl}\delta_{km})
$$
There is also a similar expression for $C_{\Phi_{IJ}\Phi_{KL}\Phi_{MN}}$.
We also have
\be
C_{\Phi_{ij}X_{kI}X_{lJ}}=\delta_{IJ}(\delta_{ik}\delta_{jl}+\delta_{il}\delta_{jk})
\ee
\be
C_{\tilde\Phi_{IJ}X_{iK}X_{jL}}=\delta_{ij}(\delta_{IK}\delta_{JL}+\delta_{IL}\delta_{JK})
\ee
all others being zero.
Consider now the perturbation
\be
\delta S=\int (f_{ij}\Phi_{ij}+\tilde f_{IJ}\tilde \Phi_{IJ}+g_{iI}X_{iI})
\ee
where $f_{ij},\tilde f_{ij}, g_{iI}$ are classically marginal coupling constants.
The relevant flow equations are\footnote{We rescale the couplings in order to absorb the $2\pi^{d\over 2}$ factors.}
\be
\dot f_{ij}=-8(f^2)_{ij}-2(gg^T)_{ij}\sp \dot{\tilde f}_{IJ}=-8(\tilde f^2)_{IJ}-2(g^Tg)_{IJ}
\sp \dot g_{iI}=-2(g\tilde f)_{iI}-2(fg)_{iI}
\ee
They imply that the UV fixed point, described by $CFT_1\times CFT_2$ is completely unstable,
and flows logarithmically towards the IR.This is in agreement with a similar analysis in \cite{klebanov}.

If on the other hand we consider the operators to have dimensions other than $d/2$, then the $\beta$ functions acquire
linear terms with varying signs, and it becomes possible by tuning couplings to preserve conformal invariance.

\section{Two-dimensional large-N conformal gauge theories\label{2d}}

In this appendix we will describe an example of a large-N CFT in two dimensions. It will usefull in order to support some of our claims
concerning the coupling of two or more such CFTs.

A two dimensional large-N CFT must have $c\sim {\cal O}(N^2)$. The reason is that the stress tensor
is a single trace operator, and if we normalize its two-point function, we must obtain according to the standard large-N counting
$\langle ~T^n~\rangle_c\sim N^{2-n}$.
The normalized stress tensor is $O_T=\sqrt{2\over c}~T$, so that $\langle ~ O_T^n~\rangle_c\sim c^{1-{n\over 2}}$.
We deduce that $c\sim {\cal O}(N^2)$.

One example of a  large-N gauge-theory (CFT) can be obtained by gauging the diagonal $SU(N)_{k_1+k_2}$
global symmetry of the WZW model $SU(N)_{k_1}\times SU(N)_{k_2}$
to obtain the coset CFT
\be
CFT\equiv {SU(N)_{k_1}\times SU(N)_{k_2}\over SU(N)_{k_1+k_2}}\sp c={k_1k_2(k_1+k_2+2N)(N^2-1)\over (k_1+N)(k_2+N)(k_1+k_2+N)}
\label{3999}\ee
To take the 't Hooft limit we define
\be
\lambda_1={N\over k_1}\sp \lambda_2={N\over k_2}
\label{4099}\ee
and we take $N\to \infty$ keeping $\lambda_i$ fixed\footnote{This large-N limit is different from the one that gives rise to realizations of
$W_{\infty}$ symmetry in two dimensional CFTs \cite{bakas}. This second large-N limit is associated to pp-wave type space-times.}.
We may rewrite the central charge as
\be
c={(\l_1+\l_2+2\l_1\l_2)\over (1+\l_1)(1+\l_2)(\l_1+\l_2+\l_1\l_2)}(N^2-1)
\label{4111}\ee
which has the correct large-N asymptotics.

There is an interesting symmetry in this theory, rank-level duality, that indicates that this CFT is equivalent to a dual
one\footnote{This has been explicitly checked in the associated supersymmetric models, \cite{sch} although it is also plausible here.}
\be
CFT\sim \tilde{CFT}\equiv {SU(k_1+k_2)_N\over SU(k_1)_N\times SU(k_2)_N\times U(1)}
\label{4211}\ee
with associated 't Hooft couplings
\be
\tilde\lambda_1={1\over \l_1}={k_1\over N}\sp \tilde\lambda_2={1\over \l_2}={k_2\over N}
\label{43}\ee
In the dual theory there are two types  of colors, with multiplicities $k_1$ and $k_2$.

In the special case $k_1=k_2$ we obtain
\be
CFT\equiv {SU(N)_{k}\times SU(N)_{k}\over SU(N)_{2k}}\sp
c={2k^2(N^2-1)\over (k+N)(2k+N)}
\label{44}\ee
with 't Hooft coupling
\be
\lambda={N\over k}\sp c={2(N^2-1)\over (1+\l)(2+\l)}
\label{45}\ee
while the dual one is
\be
\tilde{CFT}\equiv {SU(2k)_N\over SU(k)_N\times SU(k)_N\times U(1)}
\label{46}\ee
with 't Hooft coupling
\be
\tilde\lambda={2\over \l}={2k\over N}
\label{47}\ee
It should be noted that the rank-level duality here inverts the 't Hooft coupling.
There is no analogous case in four-dimensions.

We now proceed to analyze  the conformal dimensions.
The conformal dimensions for the primary fields of the HW reps of the SU(N)$_k$  theory are
given by
\be
\Delta_R={C_2(R)\over k+N}={C_2(R)\over N}{\l\over \l+1}
\label{511}\ee
We must therefore analyze the scaling of the quadratic Casimir for SU(N) representations.

The quadratic Casimir invariant for SU(N) is given by
\be
(T_R^aT_R^a)_{ij}=C_2(R)\delta_{ij}
\label{48}\ee
while the second Dynkin index is
\be
Tr[T^a_RT^b_R]=S_2(R)\delta^{ab}\sp \dim(G)S_2(R)=\dim(R)C_2(R)
\label{49}\ee
Using the results from reference \cite{patera}
we can tabulate below in table 1 for various SU(N) reps, the dimension, second Dynkin index, quadratic Casimir
and asymptotic conformal dimension defined as
\be
\Delta_R={C_2(R)\over k+N}={C_2(R)\over N}{\l\over \l+1}\simeq \Delta(\infty){\l\over \l+1}+{\cal O}\left({1\over N}\right)
\label{51}\ee

\begin{table}[!b]
\begin{tabular}{|c|c|c|c|c|} \hline
Representation & dimension & Dynkin Index $S_2$ & Casimir $C_2$& $\Delta(\infty)$\\
\hline \hline

$\Yboxdim8pt\yng(1)$ & $N$ & ${1\over 2}$ & ${N^2-1\over 2N}$&${1\over 2}$\\

$\Yboxdim8pt\yng(2)$   &  ${N(N+1)\over 2}$ & ${N+2\over 2}$ & ${(N-1)(N+2)\over N}$&1\\

$\Yboxdim8pt\yng(1,1)$   &  ${N(N-1)\over 2}$ & ${N-2\over 2}$ &${(N+1)(N-2)\over N}$&1\\

Adjoint &  $N^2-1$& $N$&$N$&1\\

$\Yboxdim8pt\yng(3)$& ${N(N+1)(N+2)\over 6}$& ${(N+2)(N+3)\over 4}$&${3(N-1)(N+3)\over 2N}$ &${3\over 2}$\\

$\Yboxdim8pt\yng(2,1)$& ${N(N^2-1)\over 3}$& ${N^2-3\over 2}$&${3(N^2-3)\over 2N}$& ${3\over 2}$\\

$\Yboxdim8pt\yng(1,1,1)$& ${N(N-1)(N-2)\over 6}$& ${(N-2)(N-3)\over 4}$&${3(N+1)(N-3)\over 2N}$  &${3\over 2}$\\

$\Yboxdim8pt\yng(4)$& ${N(N+1)(N+2)(N+3)\over 24}$& ${(N+2)(N+3)(N+4)\over 12}$&${2(N-1)(N+4)\over N}$ &2\\

$\Yboxdim8pt\yng(1,1,1,1)$& ${N(N-1)(N-2)(N-3)\over 24}$& ${(N-2)(N-3)(N-4)\over 12}$&${2(N+1)(N-4)\over N}$ &2\\

$\Yboxdim8pt\yng(2,2)$& ${N^2(N^2-1)\over 12}$& ${(N+2)(N+3)(N+4)\over 12}$&${2(N^2-4)\over N}$&2 \\

$\Yboxdim8pt\yng(3,1)$& ${N(N-1)(N+1)(N+2)\over 8}$&  ${(N+2)(N^2+N-4)\over 4}$&${2(N^2+N-4)\over N}$ &2\\

$\Yboxdim8pt\yng(2,1,1)$& ${N(N+1)(N-1)(N-2)\over 8}$&  ${(N-2)(N^2-N-4)\over 4}$&${2(N^2-N-4)\over N}$&2 \\

m-symmetric & $\left(N+m-1\atop m\right)$& ${1\over 2} \left(N+m\atop m-1\right)$  &${m(N-1)(N+m)\over 2N}$&  ${m\over 2}$ \\

m-antisymmetric & $\left(N\atop m\right)$& ${1\over 2} \left(N-2\atop m-1\right)$  &${m(N-m)(N+1)\over 2N}$ & ${m\over 2}$\\
\hline
\end{tabular}
\label{table11}
\caption{Relevant data for some low-lying SU(N) representations}
\end{table}

We may now consider the holomorphic scaling dimensions of operators of the coset
${SU(N)_{k_1}\times SU(N)_{k_2}\over SU(N)_{k_1+k_2}}$, which in the simplest cases are in one-to one correspondence with
$R_1\in SU(N)_{k_1}$, $R_2\in SU(N)_{k_2}$, $ R_1\otimes R_2\sim R\in SU(N)_{k_1+k_2}$.
We obtain
\be
\Delta_{\Yboxdim4pt\yng(1),\Yboxdim4pt\yng(1);R}={\l_1^2+\l_2^2+\l_1\l_2(\l_1+\l_2)\over
(1+\l_1)(1+\l_2)(\l_1+\l_2+\l_1\l_2)}+{\cal O}\left({1\over N}\right)\sp R=\Yboxdim6pt\yng(1,1),\Yboxdim6pt\yng(2),{\rm adjoint}
\label{52}\ee
\be
\Delta_{\Yboxdim4pt\yng(1),1;\Yboxdim4pt\yng(1)}={1\over 2}{\l_1^2\over (1+\l_1)(\l_1+\l_2+\l_1\l_2)}+{\cal O}\left({1\over N}\right)
\label{53}\ee

\be
\Delta_{R,1;R}={\l_1^2\over (1+\l_1)(\l_1+\l_2+\l_1\l_2)}+{\cal O}\left({1\over N}\right)\sp
R=\Yboxdim6pt\yng(1,1),\Yboxdim6pt\yng(2),{\rm adjoint}
\label{54}\ee

\be
\Delta_{\Yboxdim4pt\yng(1),\overline{\Yboxdim4pt\yng(1)};1}={1\over 2}\left[{\l_1\over (1+\l_1)}+{\l_2\over (1+\l_2)}\right]+{\cal O}\left({1\over N}\right)
\label{55}\ee

We will now investigate which of the coset fields correspond to single trace operators.
To do this we must start from the fundamental fields $g_{1,2}$ of the WZW theories $SU(N)_{k_i}$
transforming in the ($\Yboxdim4pt\yng(1),\overline{\Yboxdim4pt\yng(1)}$) representation of the $SU(N)_{L}\times SU(N)_R$
global symmetry.
Under the U(N) gauge symmetry of the coset theory, they transform as
\be
g_1\to \bar U ~g_1~ V\sp g_2\to \bar U ~g_2~ V
\label{56}\ee
Then the operator $Tr[g_1^{-1}g_2]$ is gauge invariant and therefore a valid coset primary field, which moreover is a single-trace operator.
In fact, if one takes into account the OPE product expansion, there are two coset primaries, associated with $Tr[g_1^{-1}g_2]$, namely
($\Yboxdim4pt\yng(1),\overline{\Yboxdim4pt\yng(1)}~;1)$ and ($\Yboxdim4pt\yng(1),\overline{\Yboxdim4pt\yng(1)}~;{\rm adjoint})$
with large-N dimensions
\be
\Delta_{\Yboxdim4pt\yng(1),\overline{\Yboxdim4pt\yng(1)};1}={1\over 2}\left[{\l_1\over (1+\l_1)}+{\l_2\over (1+\l_2)}\right]
\sp \Delta_{\Yboxdim4pt\yng(1),\overline{\Yboxdim4pt\yng(1)}~;{\rm adjoint}}={\l_1^2+\l_2^2+\l_1\l_2(\l_1+\l_2)\over
(1+\l_1)(1+\l_2)(\l_1+\l_2+\l_1\l_2)}
\label{57}\ee

A similar argument indicates that the primary fields associated with ($R,\bar R,X$) with $X$ appearing in $R\otimes \bar R$
correspond to single trace operators.
We obtain for example
\be
\Delta_{\Yboxdim4pt\yng(2),\overline{\Yboxdim4pt\yng(2)};1}=\left[{\l_1\over (1+\l_1)}+{\l_2\over (1+\l_2)}\right]
\label{58}\ee
\be
 \Delta_{\Yboxdim4pt\yng(2),\overline{\Yboxdim4pt\yng(2)}~;{\rm adjoint}}={\l_1^2+\l_2^2+\l_1\l_2+\l_1\l_2(2\l_1+2\l_2+\l_1\l_2)\over
(1+\l_1)(1+\l_2)(\l_1+\l_2+\l_1\l_2)}
\label{59}\ee


\end{document}